\documentclass[iop]{emulateapj}

\newcommand{\vycn}{C$_2$H$_3$CN}
\newcommand{\etcn}{C$_2$H$_5$CN}
\newcommand{\kms}{\mbox{km\ \ensuremath{\rm{s}^{-1}}}}

\newcommand{\eg}{\emph{e.g.}}

\slugcomment{Accepted for publication in the Astronomical Journal}

\shorttitle{Mapping Vinyl Cyanide}
\shortauthors{Lai. Cordiner, et al.}

\begin{document}

\title{Mapping Vinyl Cyanide and Other Nitriles in Titan's Atmosphere Using ALMA}

\author{J. C.-Y. Lai \altaffilmark{1,2}, M. A. Cordiner\altaffilmark{1,3}, C. A. Nixon\altaffilmark{1}, R. K. Achterberg\altaffilmark{1},  E. M. Molter\altaffilmark{1,4}, N. A. Teanby\altaffilmark{5}, M. Y. Palmer\altaffilmark{1,3}, S. B. Charnley\altaffilmark{1}, J. E. Lindberg\altaffilmark{1}, Z. Kisiel\altaffilmark{6}, M. J. Mumma\altaffilmark{1}, P. G. J. Irwin\altaffilmark{7}}

\altaffiltext{1}{NASA Goddard Space Flight Center, 8800 Greenbelt Road, Greenbelt, MD 20771, USA}
\email{martin.cordiner@nasa.gov}
\altaffiltext{2}{McMaster University, Hamilton, Ontario, L8S 4L8, Canada}
\altaffiltext{3}{Department of Physics, Catholic University of America, Washington, DC 20064, USA}
\altaffiltext{4}{Astronomy Department, 501 Campbell Hall, University of California, Berkeley, CA 94720, USA}
\altaffiltext{5}{School of Earth Sciences, University of Bristol, Wills Memorial Building, Queen's Road, Bristol, BS8 1RJ, UK}
\altaffiltext{6}{Institute of Physics, Polish Academy of Sciences, Al. Lotnik{\o}w 32/46, 02-668 Warszawa, Poland}
\altaffiltext{7}{Atmospheric, Oceanic and Planetary Physics, Clarendon Laboratory, University of Oxford, Parks Road, Oxford, OX1 3PU, UK}

\begin{abstract}
Vinyl cyanide (\vycn) is theorized to form in Titan's atmosphere via high-altitude photochemistry and is of interest regarding the astrobiology of cold planetary surfaces due to its predicted ability to form cell membrane-like structures (azotosomes) in liquid methane. In this work, we follow up on the initial spectroscopic detection of \vycn\ on Titan by \citet{pal17} with the detection of three new \vycn\ rotational emission lines at submillimeter frequencies. These new, high-resolution detections have allowed for the first spatial distribution mapping of \vycn\ on Titan. We present simultaneous observations of C$_2$H$_5$CN, HC$_3$N, and CH$_3$CN emission, and obtain the first (tentative) detection of C$_3$H$_8$ (propane) at radio wavelengths. We present disk-averaged vertical abundance profiles, two-dimensional spatial maps, and latitudinal flux profiles for the observed nitriles. Similarly to HC$_3$N and C$_2$H$_5$CN, which are theorized to be short-lived in Titan's atmosphere, C$_2$H$_3$CN is most abundant over the southern (winter) pole, whereas the longer-lived CH$_3$CN is more concentrated in the north. This abundance pattern is consistent with the combined effects of high-altitude photochemical production, poleward advection, and the subsequent reversal of Titan's atmospheric circulation system following the recent transition from northern to southern winter. We confirm that \vycn\ and \etcn\ are most abundant at altitudes above 200~km. Using a 300~km step model, the average abundance of \vycn\ is found to be $3.03\pm0.29$~ppb, with a C$_2$H$_5$CN/\vycn\ abundance ratio of $2.43\pm0.26$. Our HC$_3$N and CH$_3$CN spectra can be accurately modeled using abundance gradients above the tropopause, with fractional scale-heights of $2.05\pm0.16$ and $1.63\pm0.02$, respectively.
\end{abstract}


\section{Introduction}

Titan, Saturn's largest moon, possesses the most substantial atmosphere of any known satellite in the solar system. This atmosphere has been shown to contain a variety of organic compounds, including hydrocarbons as well as nitrogen- and oxygen-containing species \citep[see the review by][]{bez14}. Due to their often-large dipole moments and strong rotational transitions, nitriles (molecules containing a CN group) are readily detected in the millimeter/submillimeter range. By virtue of the rapid production in the upper atmosphere and relatively short chemical lifetimes of some nitriles \citep{wil04}, these molecules can be used as powerful probes of the seasonally variable chemistry and dynamics in Titan's upper atmosphere \citep[\eg][]{tea12, cor14, vin15}.  

Vinyl cyanide, or acrylonitrile (\vycn), has come into focus due to the recent work of \citet{ste15}, which showed, through theoretical liquid-phase calculations, that this molecule is one of the most favored to form thermodynamically stable membranes (azotosomes) in liquid methane at the surface temperature of Titan (approximately 94~K). Given that Titan possesses seas of liquid hydrocarbon, vinyl cyanide could be a strong candidate for forming membranes of potential astrobiological importance. Although the presence of \vycn\ on Titan was previously inferred from mass-spectrometric measurements of its protonated form and neutral dissociation fragments \citep{vui07,cui09}, the first definitive detection of the molecule itself was by \citet{pal17}, who reported observations of three mm-wave emission lines, making use of archival data obtained using the band 6 receiver of the Atacama Large Millimeter/submillimeter Array (ALMA) in 2014.

The \citet{pal17} observations were obtained at relatively low signal-to-noise and with insufficient resolution to provide reliable maps of the \vycn\ emission, so much remains unknown about the distribution of this important molecule within Titan's atmosphere. In this paper, we present new higher-sensitivity sub-mm data obtained from the ALMA Science Archive. Three separate vinyl cyanide lines were detected in Band 7, strengthening our confidence in the previous detections and abundance estimates. Furthermore, the high spatial resolution of these data allows us to analyze the distribution of \vycn\ across Titan for the first time. We also present new data for the 2015 April epoch on the spatial distributions of the previously detected nitrile species ethyl cyanide (C$_2$H$_5$CN), cyanoacetylene (HC$_3$N), and acetonitrile (CH$_3$CN).

\section{Observations}
\label{sec:obs}

Routine interferometric observations of Titan were performed for the purpose of flux calibration as part of ALMA project 2013.1.00033.S on 2015 April 29. A single 3-minute integration of Titan was obtained at UT 09:19 using the band 7 receiver, covering frequencies in the range 335.3-349.2~GHz. Observations were made with 39 antennas in the telescope array, which provided baselines in the range 15-349~m, with good $u-v$ (Fourier) coverage of the sky. As with the previous work of \citet{cor14,cor15} and \citet{pal17}, even this short integration time was sufficient to reach a high enough sensitivity for the detection of weak molecular emission lines from Titan. The present study was further facilitated by the larger number of functioning antennas compared to the 20-30 available in previous studies (e.g. \citet{cor14,cor15} and \citet{pal17}). The data used in our study were taken from a single spectral window in the upper receiver sideband, containing 3840 channels and with a channel spacing of 488 kHz which (after Hanning smoothing by the correlator) corresponds to a spectral resolution of 976 kHz. The low (0.57~mm) precipitable water vapor column and good phase stability provided excellent atmospheric conditions for these observations.

The data obtained from the ALMA Science Archive were processed in the NRAO CASA software (version 4.5.3) using the standard scripts provided by the Joint ALMA Observatory, as described by \citet{cor15}. The measured continuum flux density for each baseline was scaled to match the Butler-JPL-Horizons 2012 Titan flux model. To produce spectral line maps, the continuum was subtracted from the visibility amplitudes using the CASA {\tt uvcontsub} task. Imaging and deconvolution were performed using the {\tt clean} task; the point-spread function was deconvolved using the Hogbom algorithm, with natural visibility weighting and a threshold flux level of 18 mJy (twice the RMS noise level of the continuum-subtracted line-free data). The image pixel size was set to $0.1''\times0.1''$, and the spatial resolution (FWHM of the Gaussian restoring beam) was $1.22''\times0.62''$. For comparison, the angular diameter of Titan's surface was $0.78''$ on the sky. Information from the JPL Horizons system\footnote{http://ssd.jpl.nasa.gov/horizons.cgi} was used to transform the images from equatorial coordinates to projected linear distances with respect to the center of Titan and to correct the -6.90~\kms\ Doppler shift with respect to the ALMA rest frame.

\begin{figure*}
	\centering
	\includegraphics[width=0.9\textwidth]{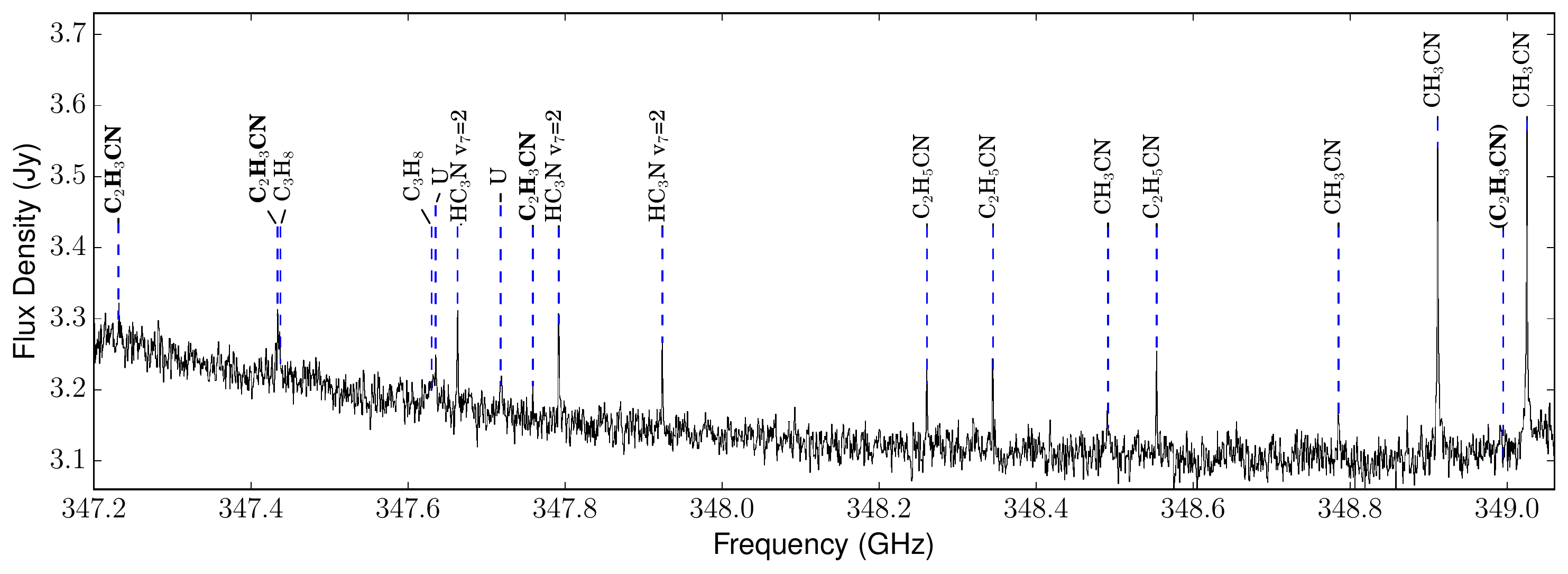}
	\caption{ALMA band 7 spectrum of Titan observed 2015 April 29. Prominent emission lines are assigned by species; bold typeface highlights the \vycn\ emission features. Two unidentified emission lines are also denoted ``U''. The upwards slope toward lower frequencies is due to pressure-broadened CO ($J=3-2$) emission. \label{fig:spectrum} }
\end{figure*}

\begin{table*}
	\centering
	\caption{Detected transitions and upper-state energies\label{tab:trans}}
	{\footnotesize
		\begin{tabular*}{\textwidth}{@{\extracolsep{\fill}}llccrr}
			\hline\hline
			Species&Line ID&Frequency (GHz)&Transition&$E_u$ (K)       &Flux (Jy\,kHz)\\
			\hline               
			C$_2$H$_3$CN     & 1& 347.232      & $38_{1,38}       $$-$$ 37_{1,37}      $ &     329&$59\pm14$\\
			C$_2$H$_3$CN     & 2& 347.434      & $38_{0,38}       $$-$$ 37_{0,37}      $ &     329&$330\pm27^a$\\
			C$_2$H$_3$CN     & 3& 347.759      & $36_{2,34}       $$-$$ 35_{2,33}      $ &     317&$41\pm12$\\
			C$_2$H$_3$CN     & 4& 348.991      & $37_{1,36}       $$-$$ 36_{1,35}      $ &     325&$60\pm15$\\[2mm]
			C$_3$H$_8$       & 1& 347.436      & $23_{0,23}       $$-$$ 22_{1,22}      $ &     202&$330\pm27^a$\\
			C$_3$H$_8$       & 2& 347.630      & $23_{1,23}       $$-$$ 22_{0,22}      $ &     202&$341\pm37^b$\\[2mm]
			HC$_3$N $v_7=2$  & 1& 347.663      & ~~~~$38           $$-$$ 37\ 0e          $ &       967&$203\pm19$\\
			HC$_3$N $v_7=2$  & 2& 347.792      & ~~~~$38           $$-$$ 37\ 2e         $ &        971&$219\pm19$\\
			HC$_3$N $v_7=2$  & 3& 347.924      & ~~~~$38           $$-$$ 37\ 2f         $ &        971&$216\pm22$\\[2mm]
			C$_2$H$_5$CN     & 1& 348.261      & $39_{2,37}       $$-$$ 38_{2,36}      $ &     344&$203\pm23$\\
			C$_2$H$_5$CN     & 2& 348.345      & $40_{2,39}       $$-$$ 39_{2,38}      $ &     351&$138\pm15$\\
			C$_2$H$_5$CN     & 3& 348.553      & $40_{1,39}       $$-$$ 39_{1,38}      $ &     351&$209\pm20$\\[2mm]
			CH$_3$CN         & 1& 348.492      & $19_{12}         $$-$$ 18_{12}        $ &     1194&$259\pm31$\\
			CH$_3$CN         & 2& 348.785      & $19_{10}         $$-$$ 19_{10}        $ &     881&$134\pm19$\\
			CH$_3$CN         & 3& 348.911      & $19_9         $$-$$ 18_9        $ &           745&$1121\pm41$\\
			CH$_3$CN         & 4& 349.025      & $19_8         $$-$$ 18_8        $ &           624&$1048\pm42$\\[2mm]
			CO               & 1& 345.796      & $3$$-$$2$               &          33.2&---\\[2mm]  
			U                & 1& 347.635      &              $-$                  &              &$330\pm27^a$\\
			U                & 2& 347.719      &              $-$                  &              &$341\pm37^b$\\
			\hline
		\end{tabular*}
	}
	\parbox{\textwidth}{\footnotesize 
		\vspace*{1mm}
		{\bf Notes.} $^a$ C$_2$H$_3$CN transition 2 and C$_3$H$_8$ transition 1 are blended.\\ $^b$ C$_3$H$_8$ transition 2 and unidentified (U) line 1 are blended.\\ Primary spectroscopic sources for molecular line frequencies: C$_2$H$_3$CN --- \citet{kis09}, C$_3$H$_8$ --- \citet{dro06}, HC$_3$N --- \citet{tho00}, C$_2$H$_5$CN --- \citet{bra09}, CH$_3$CN --- \citet{bou80}.}
\end{table*}

\begin{figure*}
	\centering
	\includegraphics[width=0.9\textwidth]{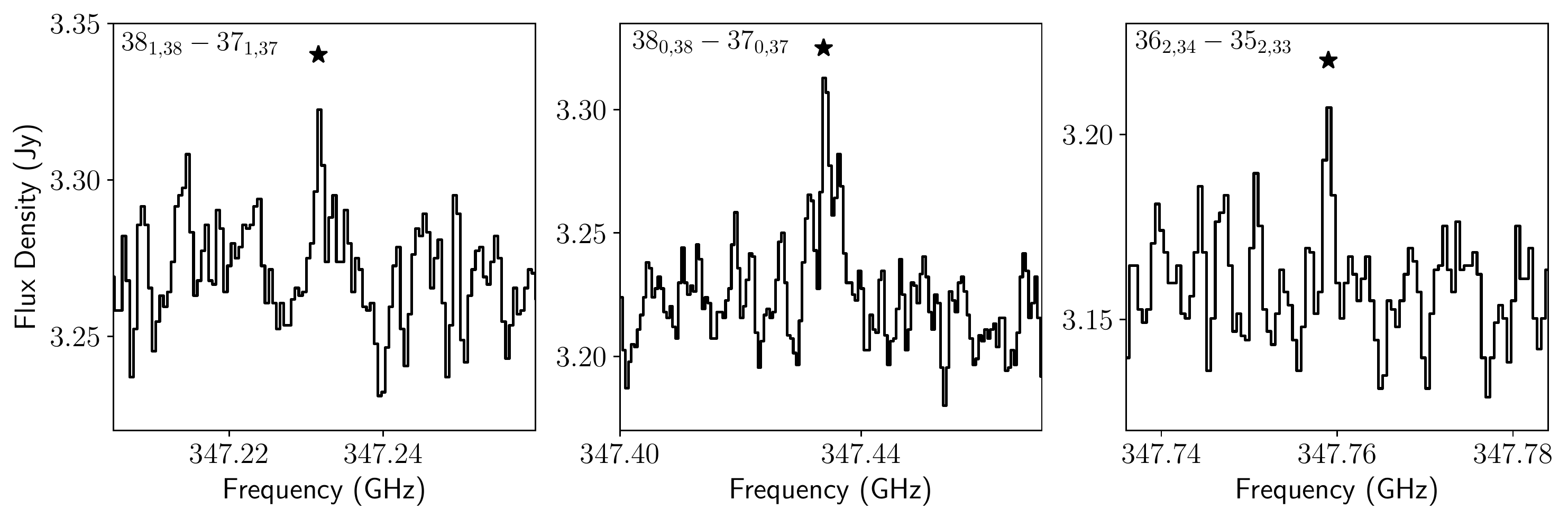}
	\caption{Close-up spectra of the three newly detected \vycn\ transitions (marked with asterisks) \label{fig:vycncloseup} }
\end{figure*}

\begin{figure*}
	\centering
	\includegraphics[width=0.45\textwidth]{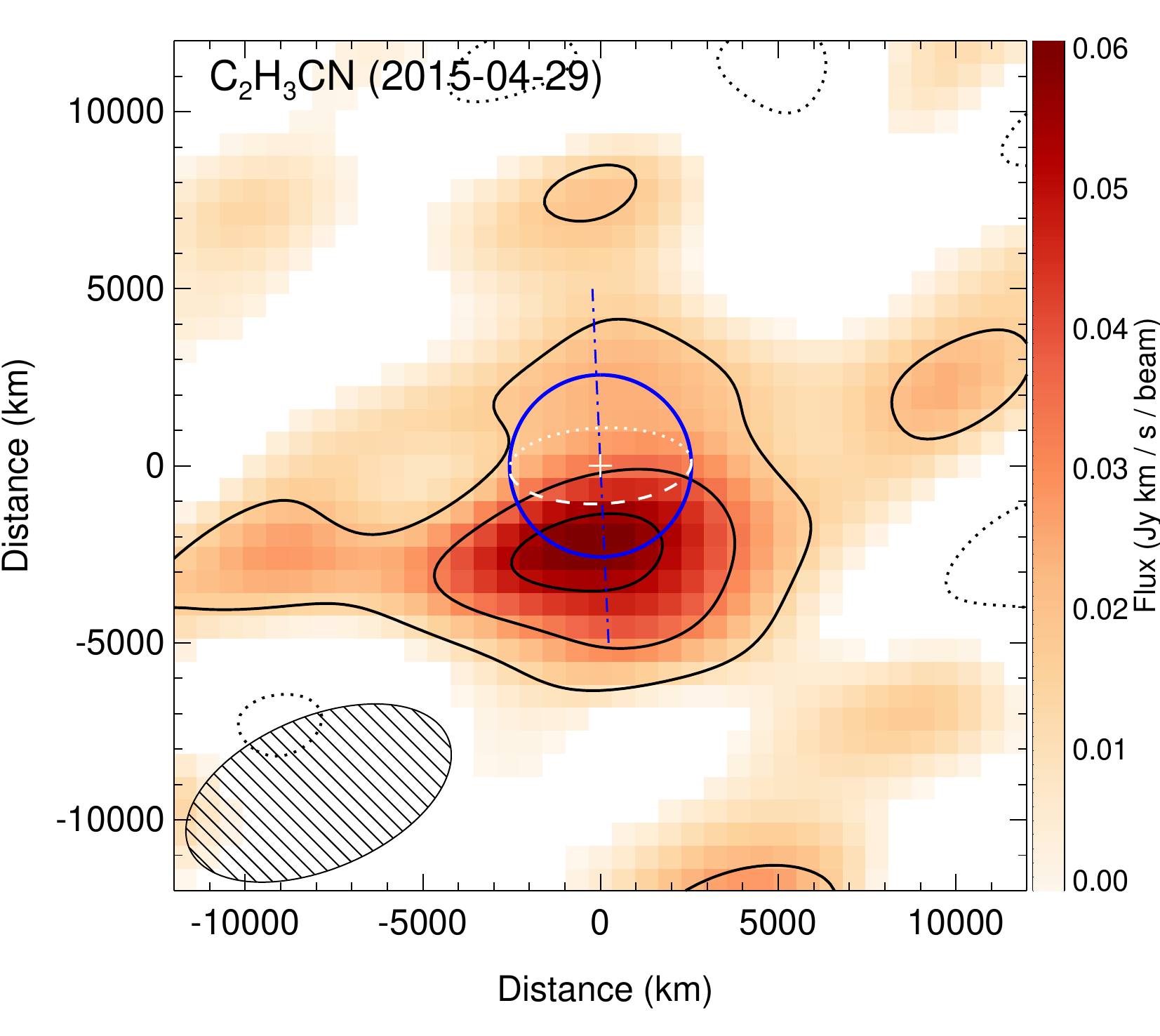}
	\hspace{10mm}
	\includegraphics[width=0.45\textwidth]{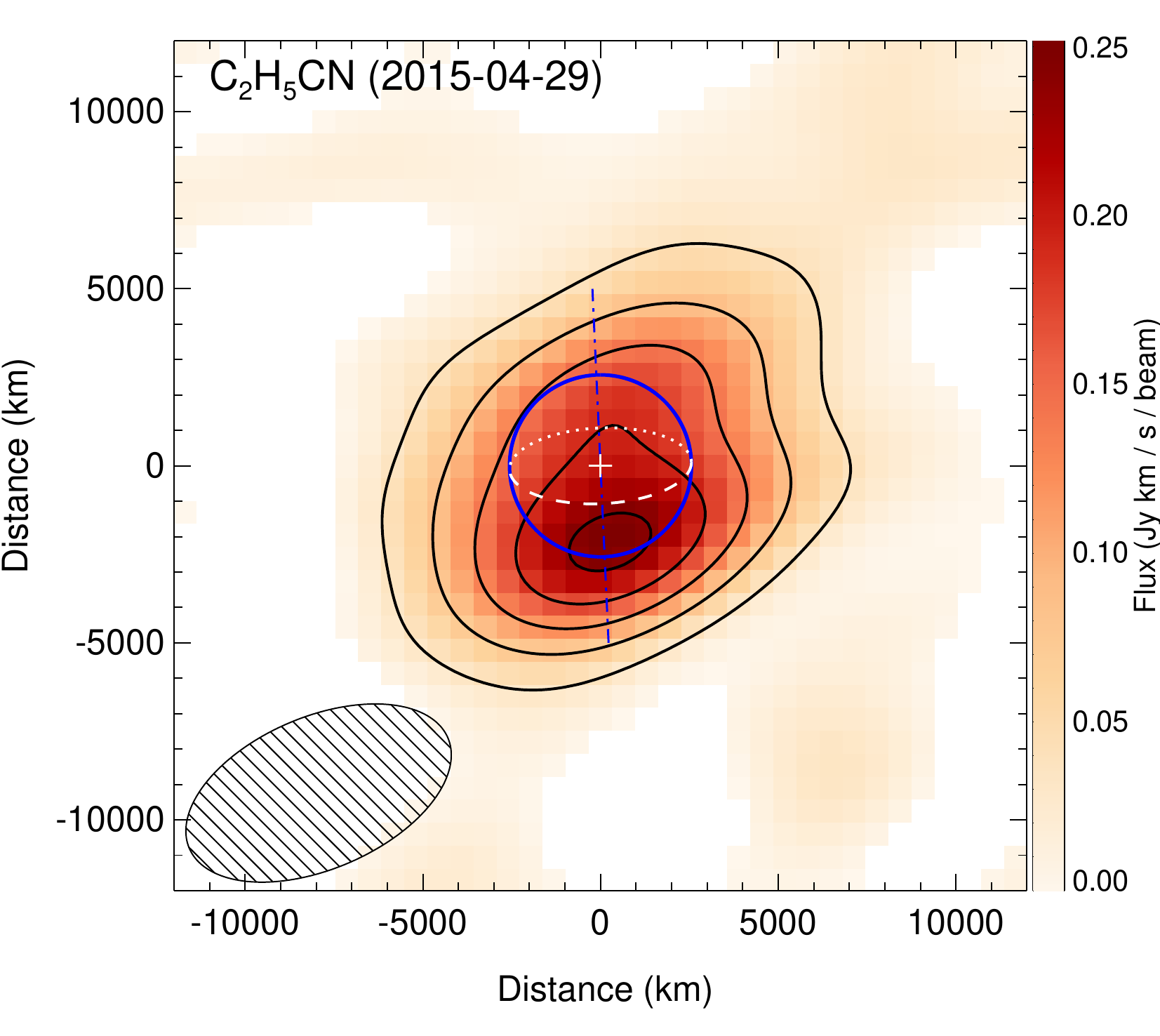}\\
	\includegraphics[width=0.45\textwidth]{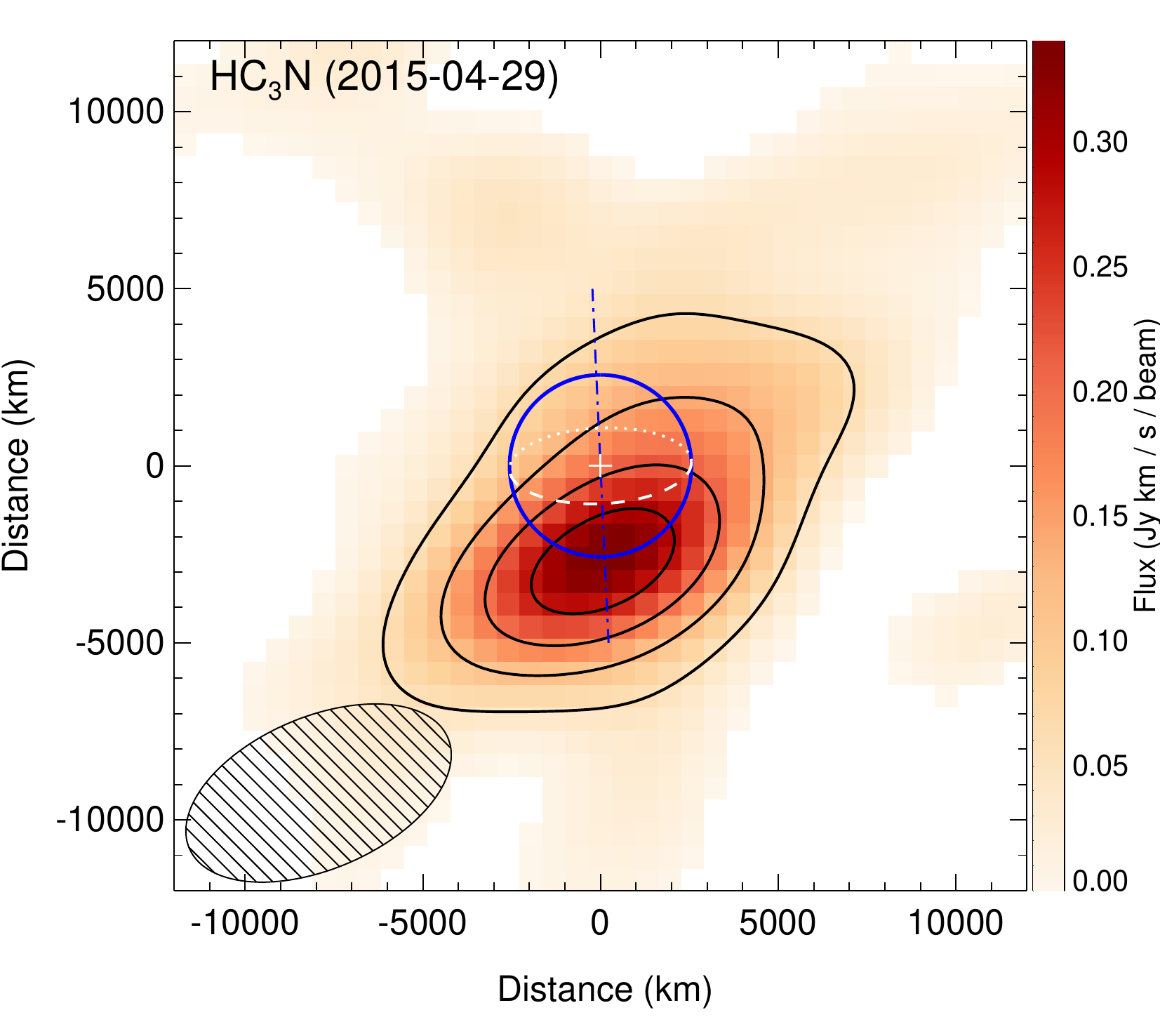}
	\hspace{10mm}
	\includegraphics[width=0.45\textwidth]{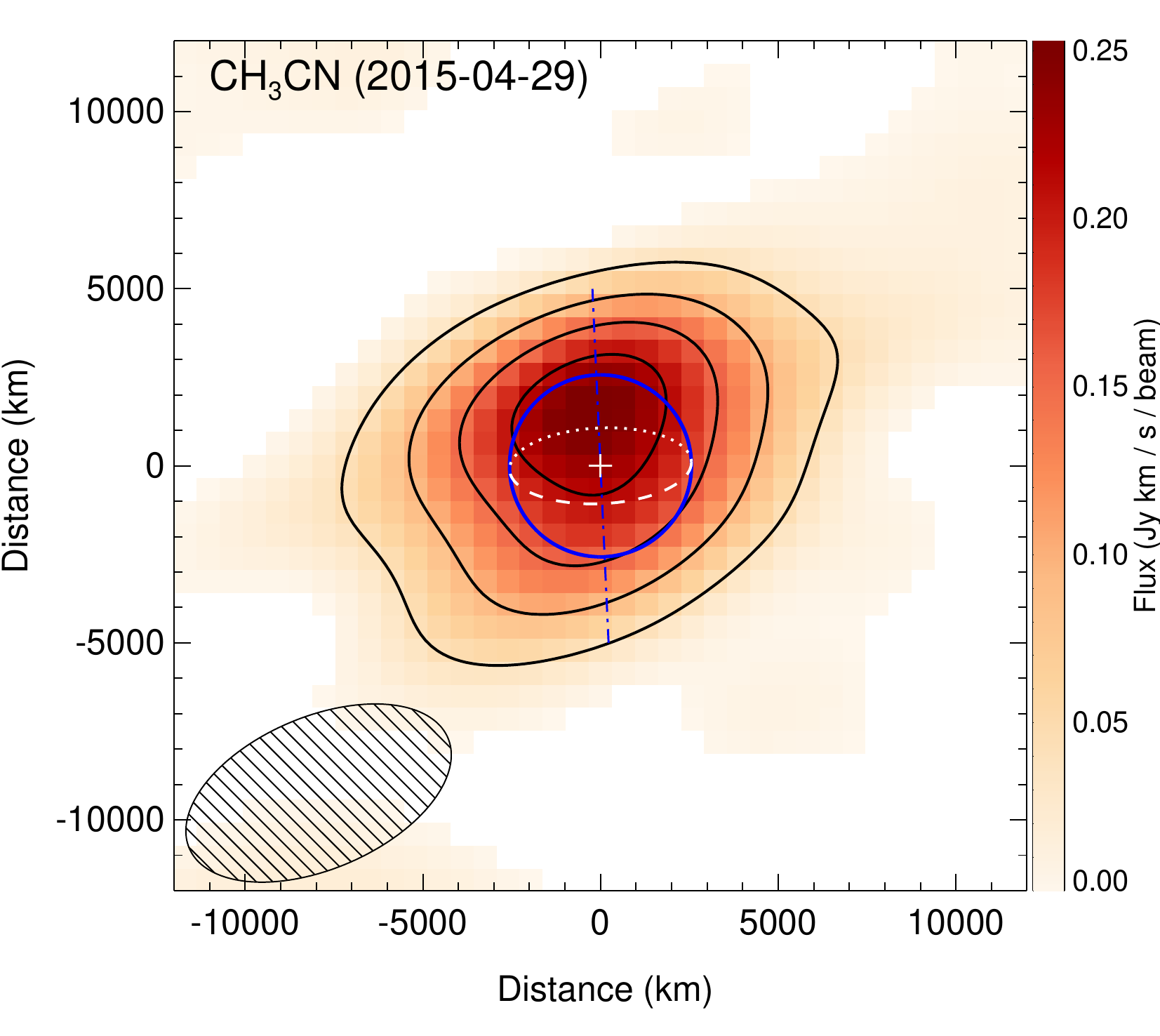}
	\caption{Integrated emission contour maps for \vycn, C$_2$H$_5$CN, HC$_3$N, and CH$_3$CN. The coordinate scale is in Titan-projected distances, and axes are aligned in the equatorial coordinate system. Titan's surface is represented by the blue circle, with the white cross denoting the position of the phase center. Dashed white curves denote Titan's equator, and a blue dotted-dashed line marks the orientation of the polar axis (2.6$^{\circ}$ counter-clockwise from vertical; north tilted toward the observer by $24.7^{\circ}$). The contour intervals (in units of $\sigma$ --- the RMS noise level of each map), are as follows: \vycn: $1.5\sigma$, C$_2$H$_5$CN: $3\sigma$, HC$_3$N: $3\sigma$, CH$_3$CN: $4\sigma$. Titan's Earth-facing hemisphere was almost fully illuminated at the time of observation, with a (Sun-target-observer) phase angle of $2.4^{\circ}$. The FWHM of the Gaussian restoring beam ($1.22''\times0.62''$) and its orientation are shown as hatched ellipses. \label{fig:maps}}
\end{figure*}

\section{Results}
\label{sec:results}

The observed spectrum is shown in Figure \ref{fig:spectrum}. This was produced by integrating over an elliptical region (with FWHM $2.3''\times1.5''$), the shape of which was determined by defining a pixel mask based on a flux threshold, which was varied until 90\% of the total continuum flux was enclosed. The value of 90\% was chosen in order to obtain the maximum signal from weak lines while excluding noisier data from the periphery of the image. The difference in line-to-continuum ratio for this region compared with a larger extraction aperture including 100\% of the detected flux was found to be negligible for the purposes of our study.  Spectral peaks were identified using the Splatalogue database for astronomical spectroscopy.\footnote{http://www.cv.nrao.edu/php/splat/} The line frequencies, quantum numbers, and upper-state energies ($E_u$) in Table \ref{tab:trans} were obtained from the Cologne Database for Molecular Spectroscopy \citep{mul01}. Three lines of \vycn\ were firmly detected ($J_{K_a,K_c} = 38_{1,38}-37_{1,37}$, $38_{0,38}-37_{0,37}$ and $36_{2,34} - 35_{2,33}$) (Figure \ref{fig:vycncloseup}), with an additional, tentative detection of the $37_{1,36}-36_{1,35}$ transition. We identified multiple lines of C$_2$H$_5$CN, CH$_3$CN, and vibrationally excited HC$_3$N in the spectrum. In addition, two lines of C$_3$H$_8$ (propane) were tentatively detected, as well as two unidentified (U) lines at 347635~MHz and 347719~MHz.

The spatial distribution of flux from each species is shown in the contour maps in Figure \ref{fig:maps}, in which the intensity of emission is indicated with an orange color scale, and the black contours correspond to increments of $n\sigma$ (where $\sigma$ is the RMS noise in each map and $n$ is a constant factor). These maps were obtained by integrating over the widths of the lines in Table \ref{tab:trans}. For \vycn, only lines 1 and 3 were used in the map due to possible contamination of lines 2 and 4 with emission from C$_3$H$_8$ and CH$_3$CN, respectively. For CH$_3$CN, only line 3 was used, to probe CH$_3$CN from a single upper-state energy level (while excluding emission from the weaker, noisier lines of this species).

\begin{figure*}
	\centering
	\includegraphics[height=0.3\textheight]{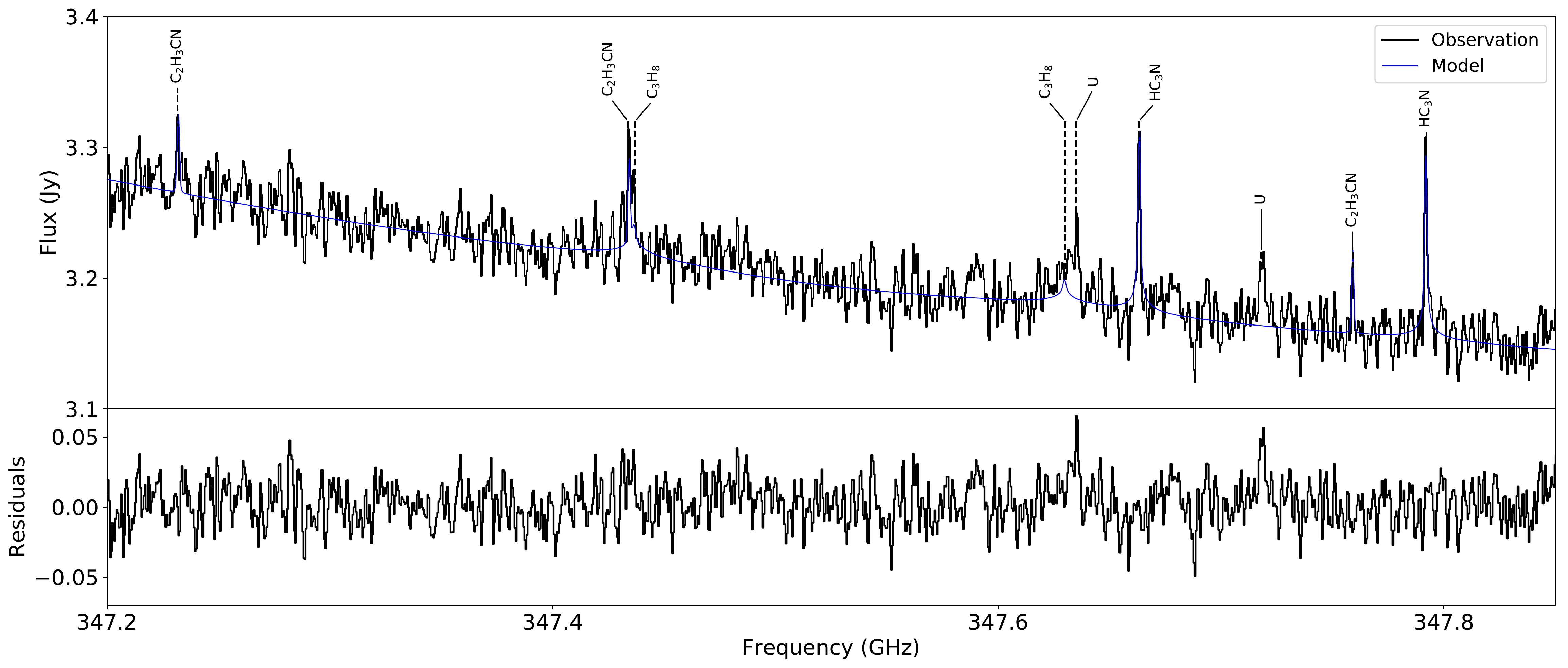}\\[2mm]
	\includegraphics[height=0.3\textheight]{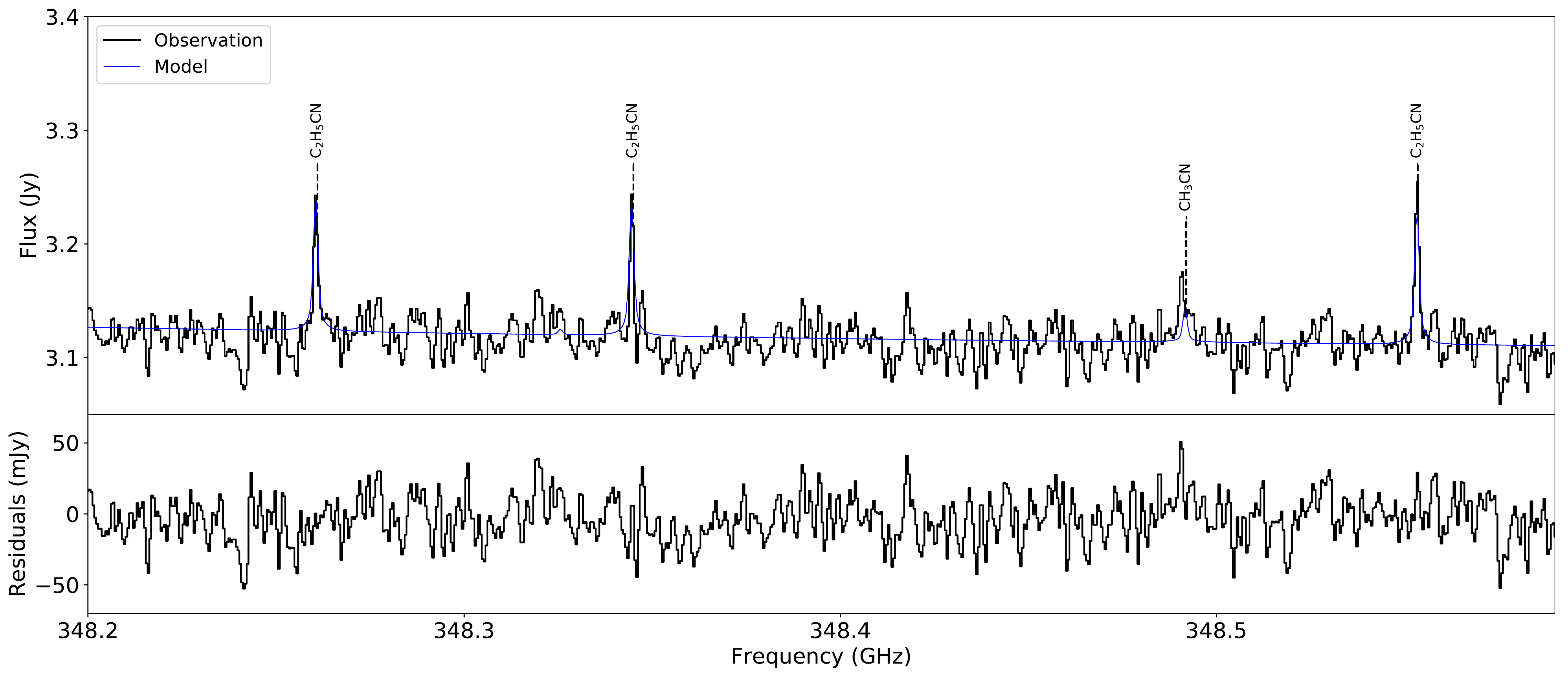}\\[2mm]
	\includegraphics[height=0.3\textheight]{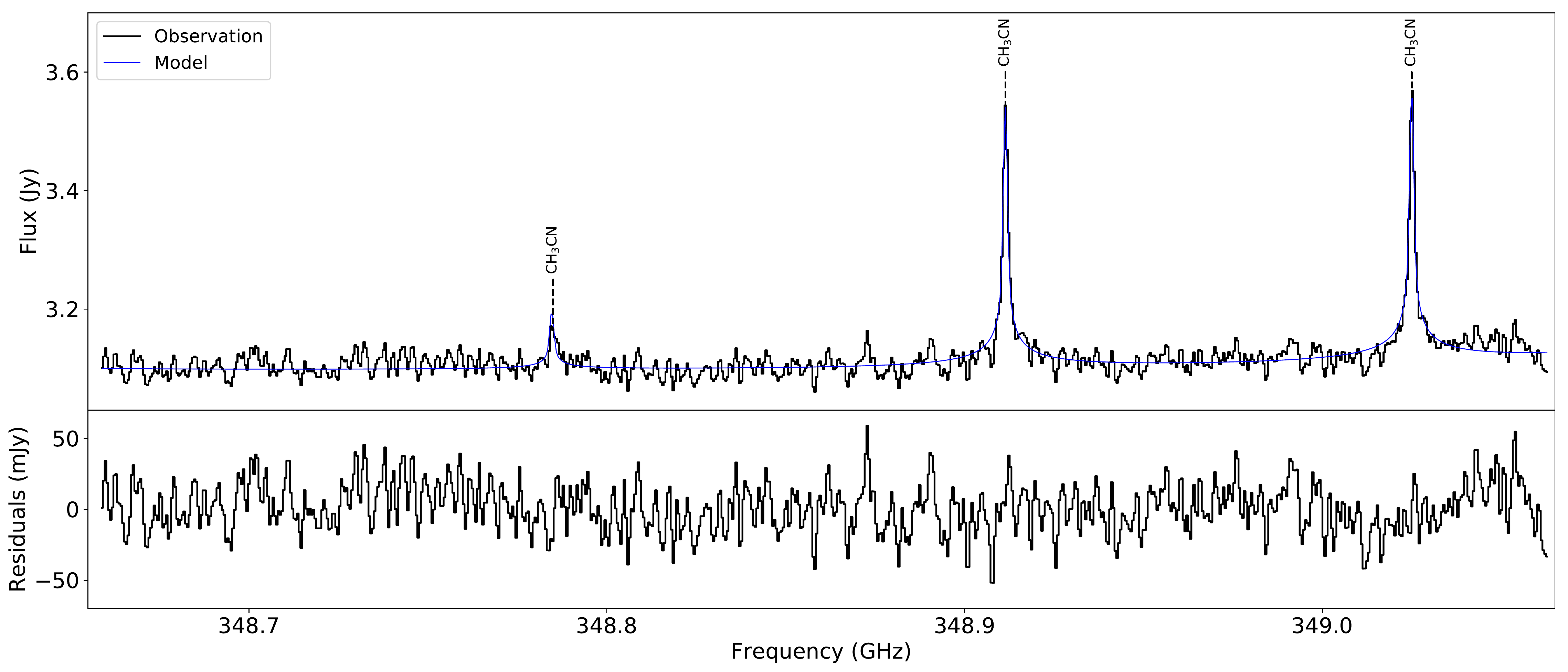}
	\caption{Observed spectra with best-fitting NEMESIS models overlaid. Lower panels show the spectroscopic residuals (observation minus model). \label{fig:models}}
\end{figure*}

\begin{figure*}
	\centering
	\includegraphics[width=0.8\textwidth]{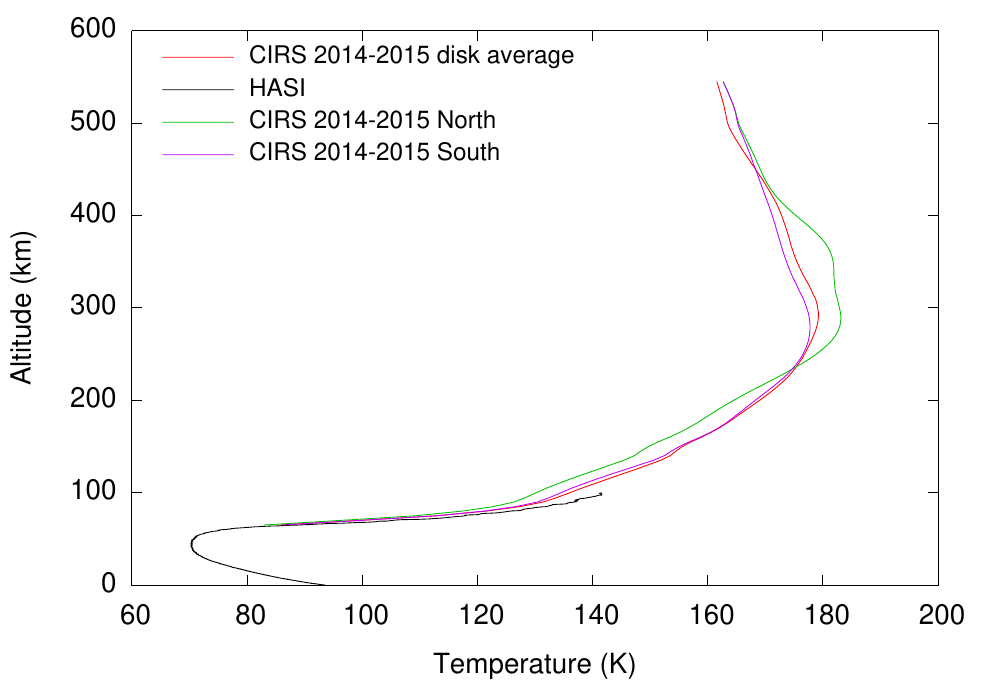}
	\caption{Temperature data used to model the spectral lines. The north and south temperature profiles are averaged over a $0.25''$ (Gaussian) beam centered over Titan's northern and southern polar limbs. \label{fig:temps}}
\end{figure*}

Following the methodology of \citet{cor14,cor15}, the observed spectrum was modeled using the line-by-line radiative transfer module of the NEMESIS atmospheric retrieval code \citep{irw08}. For computational efficiency, the data were divided into three frequency ranges: (a) 347.2-348.0~GHz, (b) 348.2-348.6~GHz, and (c) 348.6-349.1~GHz -- see Figure \ref{fig:models}. The temperature profile was based on a combination of HASI \citep{ful05} data and Cassini CIRS CH$_4$ limb-sounding data obtained during the T102, T105, T110, and T111 Titan flybys (between 2014 June and 2015 April). The CIRS temperatures as a function of latitude and altitude were retrieved following the same methodology as \citet{ach08}. The disk-averaged CIRS temperature profile (covering altitudes $z\approx60-550$~km) was combined with the HASI profile (covering $z=0$-150~km) by feathering the profiles together in the range 60-100~km using a linearly graded (weighted) average. The resulting temperature data are shown in Figure \ref{fig:temps}. As described in Section \ref{sec:maps}, latitudinal temperature variations may have a small impact on our retrieved abundances, but these are expected to be reduced to a negligible level as a result of the disk-averaging process. The abundances of nitrogen, methane, and carbon monoxide isotopologues were the same as those used by \citet{tea13}. To fit the $^{12}$CO line wing, which dominates the shape of the spectral slope in frequency range (a), the CO abundance was set to the more recent value of $5.06\times10^{-5}$ from \citet{ser16}. Propane (C$_3$H$_8$) has many transitions within our observed spectral range, but only two features (at 347.436 and 347.630~GHz) are strong enough to appear in the ALMA spectrum (and those only with a marginal significance); they were modeled using a constant C$_3$H$_8$ abundance of $4.2\times10^{-7}$ above the precipitation altitude as retrieved from CIRS observations by \citet{nix09}.

The model fluxes were calculated by integrating with a linear interpolation scheme over the radiances obtained across 35 impact parameters between the center of Titan's disk and the top of the model atmosphere at 1000~km (see \citealt{tea13} for more details). The distribution of impact parameters was varied iteratively, starting with 10 and adding more until the improvement in retrieved abundances was less than one-tenth the abundance error. The sky-projected separation between individual rays was 500~km inside Titan's disk (in the range 0-2575 km), 25~km in the range 2575-3075~km and 50~km in the range 3075-3575~km. The majority of the emission in our models originates in the altitude range 200-600~km; above 600~km, contributions are small due to exponential decay of the atmospheric density, but we extend up to 1000~km in order to avoid bias in our fractional scale-height models (described below), which decrease with altitude at a slower rate. Each ray (impact parameter) in our model was weighted by a kernel defined by the shape of the flux extraction aperture (90\% continuum flux contour) convolved with the instrumental point spread function. The observed spectra (including the pseudo-continuum of the CO line wing) were scaled up to match the model fluxes in the regions free of any other detectable line emission. This accounts for incomplete flux retrieval from the ALMA data, as well as correcting for any minor temperature differences between our model and that of the Butler-JPL-Horizons model used to calibrate the visibilities (see Section \ref{sec:obs}). 

\begin{figure*}
	\centering
	\includegraphics[width=0.745\textwidth]{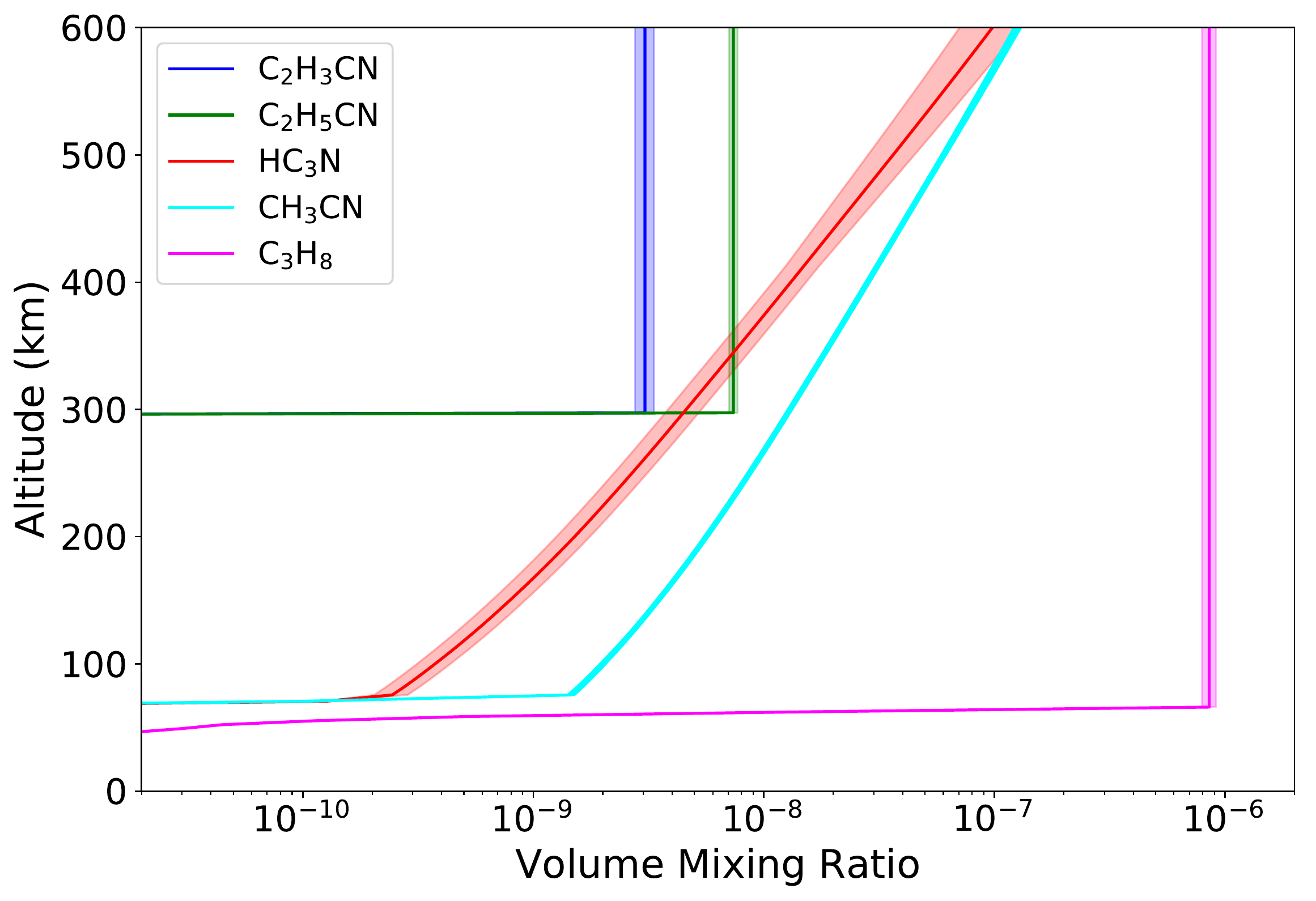}
	\caption{Vertical abundance profiles derived from radiative transfer modeling, with $1\sigma$ error envelopes. \label{fig:vmrs}}
\end{figure*}

A step model (with a constant value above 300~km and zero below) was adopted for the \vycn\ and \etcn\ vertical abundance profiles (after \citealt{cor15,pal17}), and fractional scale-height gradient models (with constant ratios of gas scale height to atmospheric scale height) were adopted for HC$_3$N and CH$_3$CN, with saturation laws taken from \citet{loi15}. A Lorentzian broadening HWHM value of $\Gamma=0.1$~cm$^{-1}$\,bar$^{-1}$, and temperature exponent $\alpha=0.75$ were used for HC$_3$N (taken from the HITRAN catalog; \citealt{gor17}). For \vycn, \etcn, and CH$_3$CN, standard values of $\Gamma=0.075$~cm$^{-1}$\,bar$^{-1}$, $\alpha=0.5$ were used due to a relative lack of detailed laboratory measurements for these species. The abundances and fractional scale-height parameters $f_H$ were optimized by minimizing the sum of squares of the spectroscopic residuals, and good fits were obtained to the spectra for all species. The spectral fits are overlaid in Figure \ref{fig:models} for each of the three frequency ranges. The retrieved abundances at 300~km altitude are listed in Table \ref{tab:models}, and the retrieved abundance profiles are plotted in Figure \ref{fig:vmrs}.

The adopted model abundance profiles were chosen to employ the least number of parameters required to obtain a reasonable fit to the observed spectra, given the noise.  As shown by \citet{cor15} and \citet{pal17}, at a similar S/N, the \vycn\ and \etcn\ 300~km step models are not unique. Furthermore, above 300~km, the line wings become negligible. Thus, the observed spectra may also be fit using `gradient' models or step-profiles with cutoffs at different altitudes. Based on the reduced $\chi^2$ values, which measure the goodness of fit, we find that models with a step height of $>$200~km provide a reasonable fit to our \vycn\ and \etcn\ observations; we adopt 300~km here to facilitate comparison with the results of \citet{pal17}. We note that our best-fitting `gradient' model has an $f_H$ value (ratio of gas scale height to atmospheric scale height) of $1.8\pm0.2$, which is somewhat less than the value of $4.6\pm1.9$ derived by \citet{pal17} and indicates a probable steepening of the \vycn\ vertical abundance profile over the time period of about a year since those observations.

\begin{table}
	\centering
	\caption{Retrieved molecular abundances at 300~km \label{tab:models}}
	\begin{tabular*}{\columnwidth}{@{\extracolsep{\fill}}lc}
		\hline\hline
		Species&Abundance (ppb)\\
		\hline
		C$_2$H$_3$CN&3.03 (0.29)\\
		C$_2$H$_5$CN&7.37 (0.32)\\
		HC$_3$N$^a$     &4.50 (0.75)\\
		CH$_3$CN$^a$    &12.7 (0.3)\\
        \hline
	\end{tabular*}
	\\
	\parbox{\columnwidth}{\footnotesize 
		\vspace*{1mm}
        NOTE --- Values in parentheses denote $1\sigma$ error margins.
		$^a$Fractional scale-heights ($f_H$) for the best-fitting HC$_3$N and CH$_3$CN `gradient' models are $2.05\pm0.16$ and $1.63\pm0.02$, respectively.\\
	}
\end{table}

\begin{figure}
	\centering
	\includegraphics[width=\columnwidth]{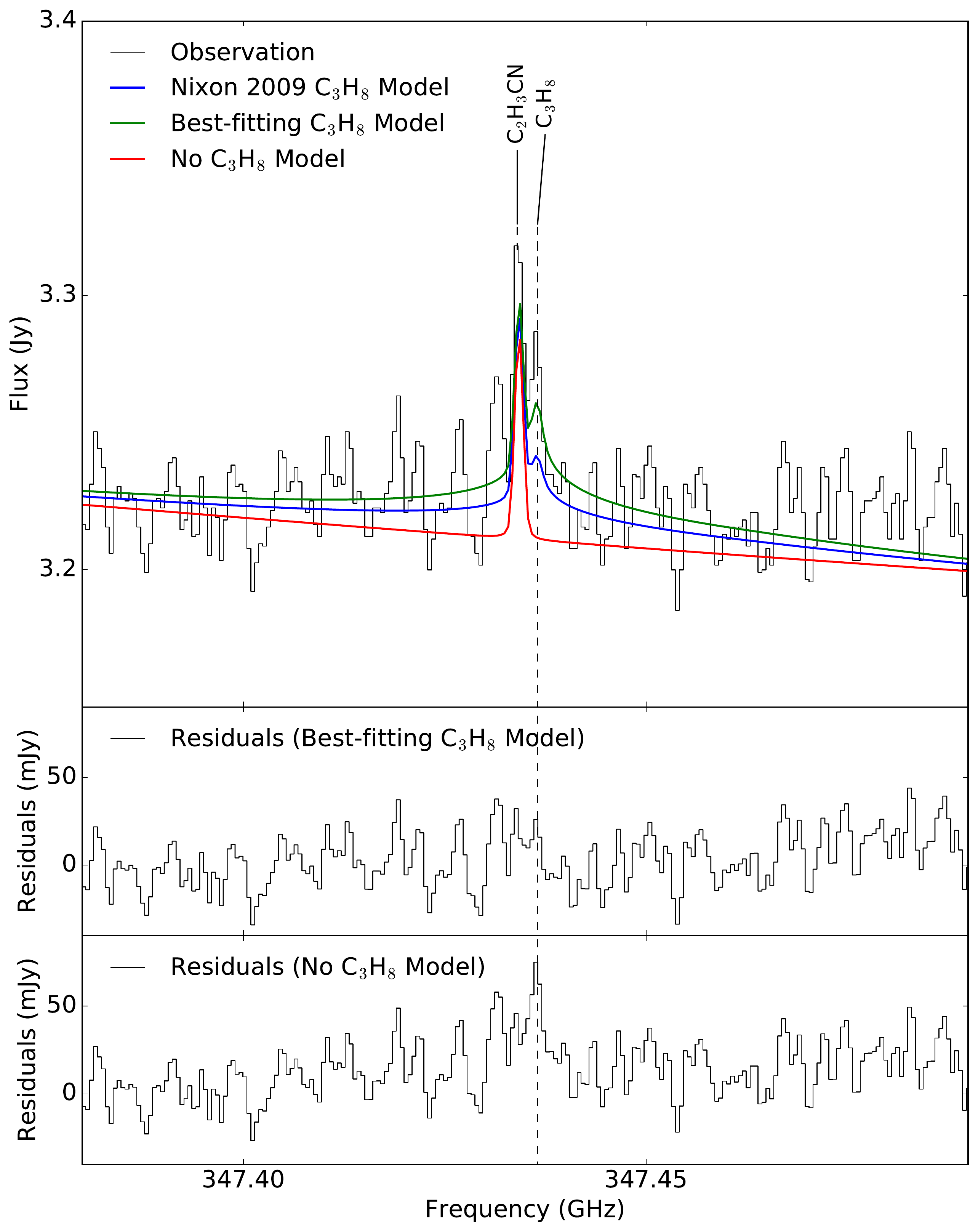}
	\includegraphics[width=\columnwidth]{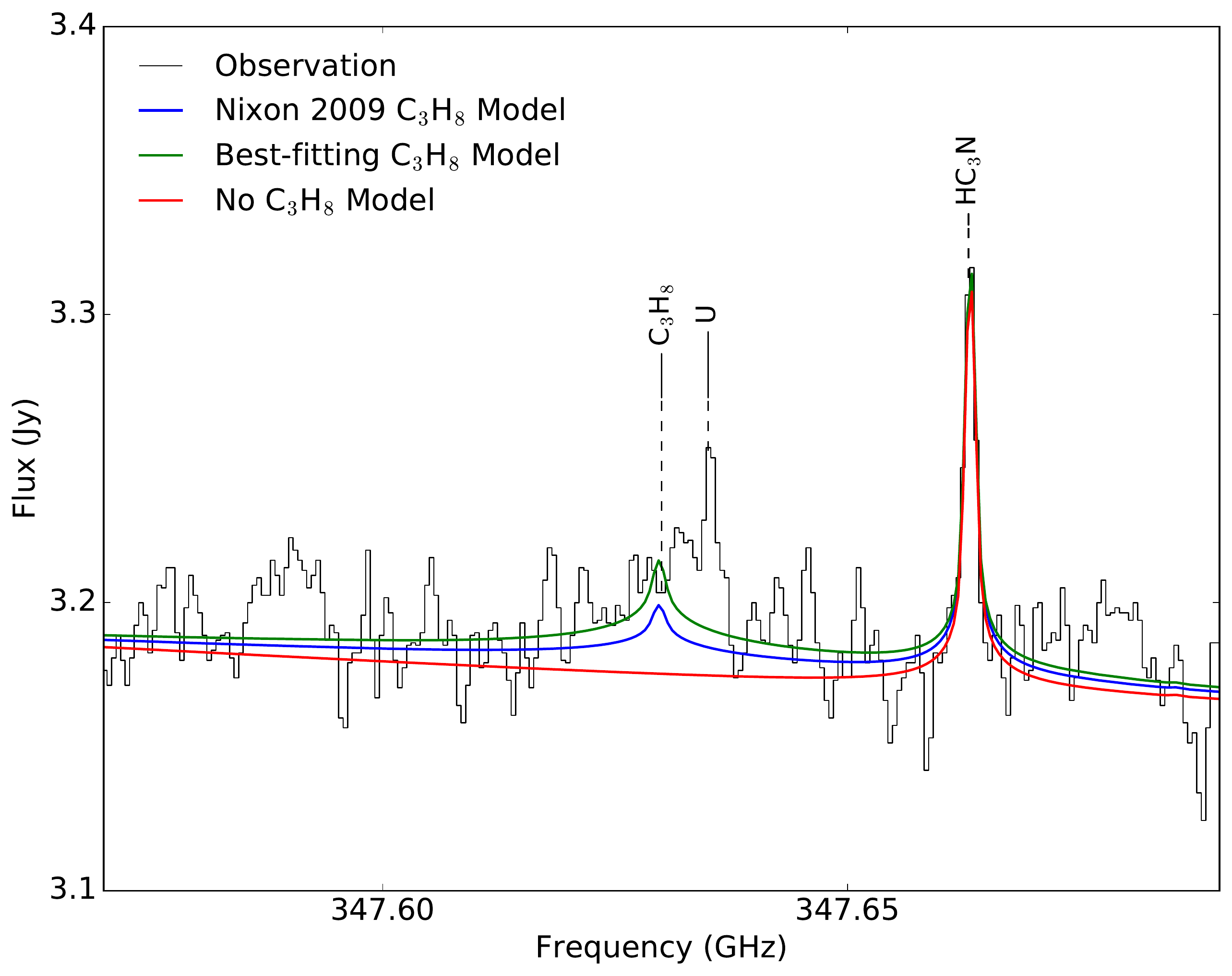}\
	\caption{Close-up view of the spectral regions surrounding the two strongest C$_3$H$_8$ transitions in our data. Three different model curves are overlaid (see the text), showing that C$_3$H$_8$ is likely to be present in our data, albeit blended with other transitions, which preclude a robust abundance estimate for this molecule. In particular, for the feature at 347.436~GHz, the probabilities that Gaussian (thermal) noise is be responsible for the observed discrepancy between the model and data are as follows: Best-fitting C$_3$H$_8$ model --- 0.30; \citet{nix09} C$_3$H$_8$ model --- 0.13; no C$_3$H$_8$ model --- 1.8$\times10^-5$. Residuals (observation minus model) are also plotted for the $23_{0,23} - 22_{1,22}$ line for the model with no C$_3$H$_8$, and the best-fitting model. Unfortunately, blending with the unidentified feature at 347.635~GHz precludes a similar statistical analysis of the feature at 347.630~GHz. However, the size of the feature predicted by the model is consistent with the observations.\label{fig:c3h8}}
\end{figure}

\begin{table}
	\centering
	\caption{Parameters of Gaussian fits to spatial profiles  \label{tab:gauss}}
	\begin{tabular*}{\columnwidth}{@{\extracolsep{\fill}}lllc}
		\hline\hline
		Species&S pos.&N pos.&Flux Ratio\\
	           &(km)  &(km)  & (South/North) 	\\
        \hline
		C$_2$H$_3$CN&-2569 (302)&2842 (770)&2.83 (0.77)\\
		C$_2$H$_5$CN&-2243 (115)&2144 (191)&1.52 (0.10)\\
		HC$_3$N     &-2790 (94) &1629 (292)&3.07 (0.29)\\
		CH$_3$CN    &-1930 (98) &2096 (60) &0.67 (0.02)\\
        \hline
	\end{tabular*}
	\\
	\parbox{\columnwidth}{\footnotesize 
		\vspace*{1mm}
        NOTE --- Distances represent the Gaussian peak positions along Titan's sky-projected polar axis (from south to north), with the sub-observer point as the origin. Values in parentheses denote $1\sigma$ error margins. Gaussian FWHM were fixed at the instrumental resolution of 4600~km.
	}
\end{table}

Vibrational emission from propane (C$_3$H$_8$) was first detected in Titan's atmosphere by \emph{Voyager 1} \citep{mag81}. Due to its theorized long photochemical lifetime ($\sim300$ years; \citealt{wil04}), propane may be expected to be quite well mixed throughout Titan's atmosphere. It shows relatively little spatial variation compared with other species (e.g. \citealt{vin15}), so we approximate the C$_3$H$_8$ distribution in our radiative transfer model using a vertical profile with a constant abundance above the saturation altitude of 60~km. Pressure broadening coefficients of $\Gamma=0.12$~cm$^{-1}$\,bar$^{-1}$, $\alpha=0.5$ were adopted based on the typical values for rovibrational transitions of C$_3$H$_8$ (see \citealt{gor17}). Allowing the C$_3$H$_8$ abundance to vary as a free parameter in our model fits, we obtained a best-fitting C$_3$H$_8$ mixing ratio of $(8.5\pm0.6)\times10^{-7}$ based on the two strongest rotational lines of this molecule in our observed spectral region. Although this is about a factor of two greater than the value of $(4.2\pm0.5)\times10^{-7}$ reported by \citet{nix09} using Cassini CIRS observations, our fitted value should be treated with extreme caution due to the fact that the two propane lines are both blended with other spectral features, leading to uncertainty in their observed strengths and profiles. The spectral regions surrounding the transitions are shown in Figure \ref{fig:c3h8}; the $23_{0,23} - 22_{1,22}$ line lies on the high-frequency side of a \vycn\ line, and the $23_{1,23}- 22_{0,22}$ line appears to be blended with the wing of an unidentified spectral feature at 347.635~GHz. In Figure \ref{fig:c3h8}, we overlay (1) the results of our best-fitting C$_3$H$_8$ radiative transfer model, (2) the predicted spectrum using the abundance obtained by \citet{nix09}, and (3) a model with no C$_3$H$_8$. We find that models including C$_3$H$_8$ provide a good fit to the observations of the $23_{0,23}- 22_{1,22}$ line whereas models without any C$_3$H$_8$ do not, as evidenced by better reduced $\chi^2$ values (1.09 with the abundance from \citet{nix09} and 1.06 with the best-fitting abundance, versus 1.73 without), which confirms the likely presence of this species in our data. However, due to the line-blending issues, we are unable to place firm constraints on the C$_3$H$_8$ abundance or its vertical profile. It should also be noted that comparison of our retrieved abundance to that of \citet{nix09} may be complicated by the fact that the value reported by \citet{nix09} represents an average over latitudes $-30^{\circ}$ to $30^{\circ}$ and does not include the polar regions. The CIRS and ALMA C$_3$H$_8$ observations are also likely to be sensitive to emission from different altitudes, due to the greatly differing spectral resolutions of these instruments.

\begin{figure*}
	\centering
	\includegraphics[width=0.45\textwidth]{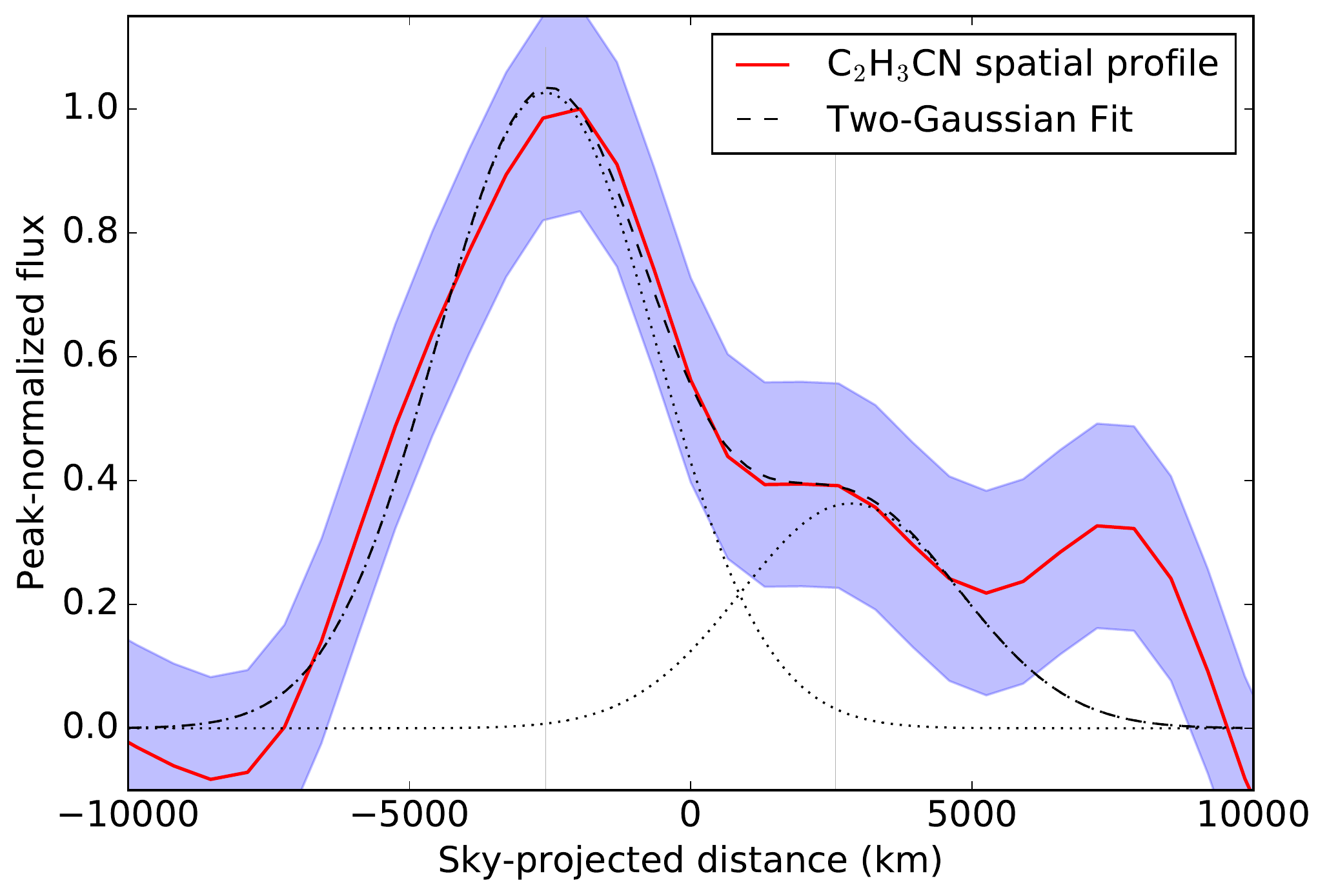}
	\includegraphics[width=0.45\textwidth]{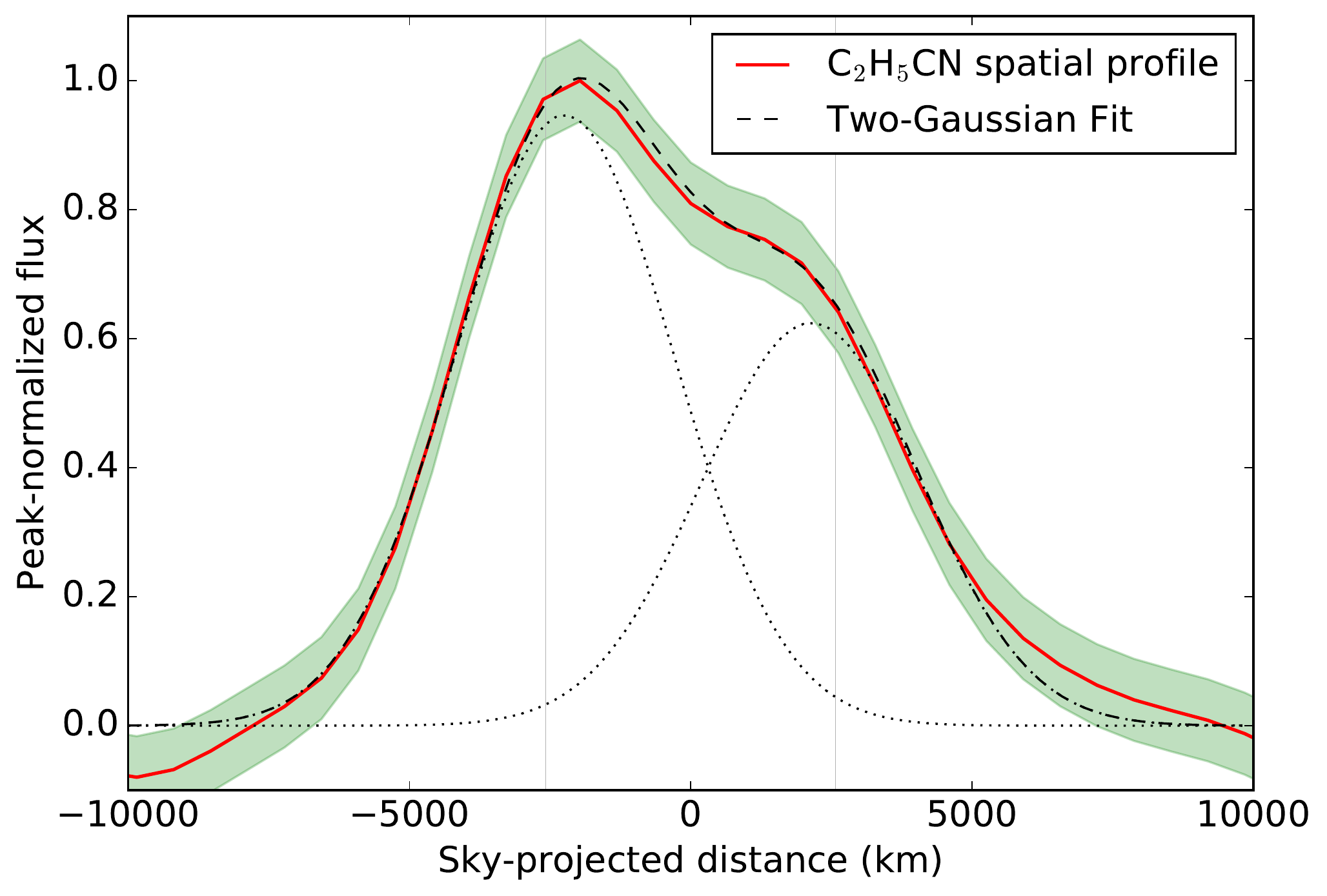}\\
	\includegraphics[width=0.45\textwidth]{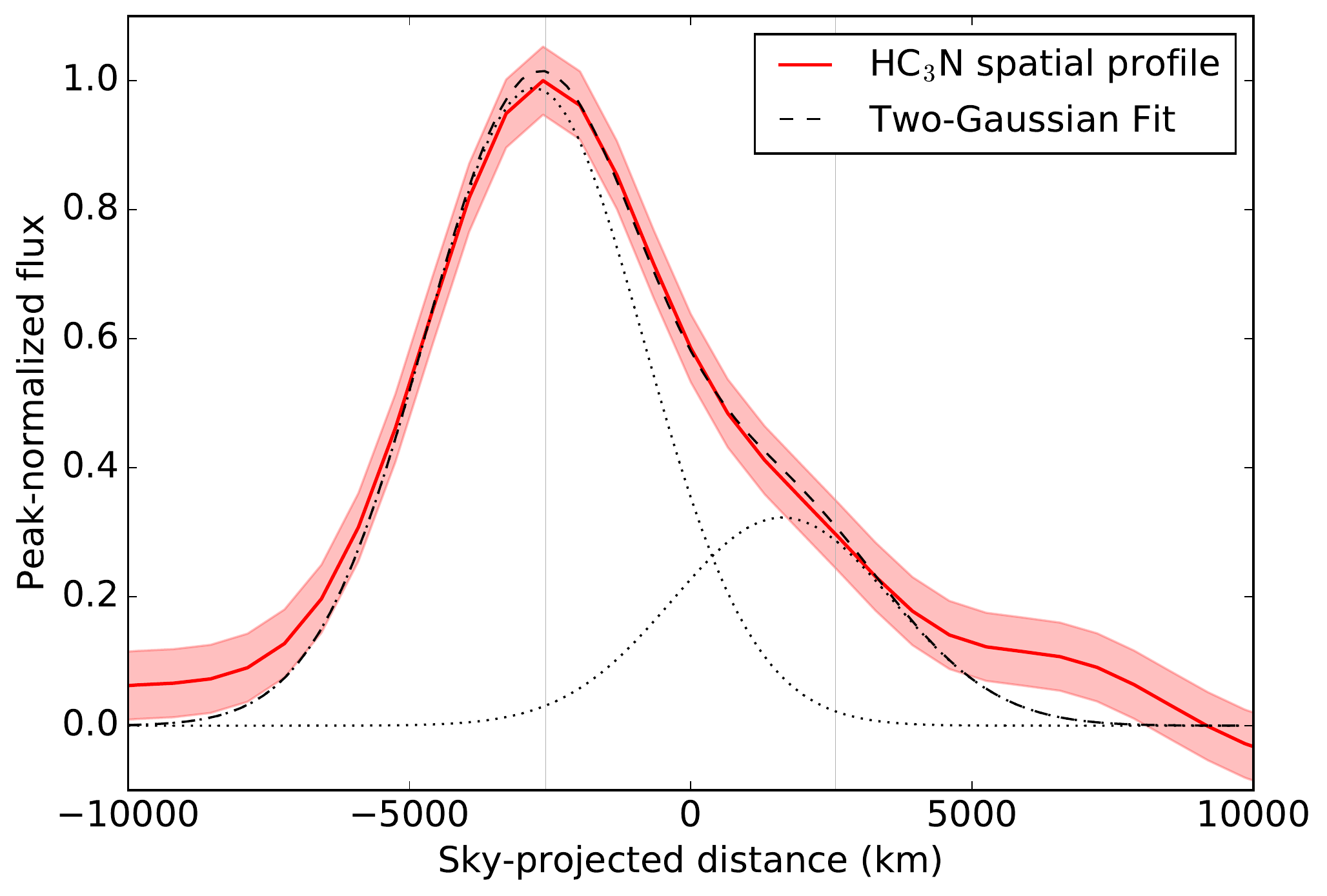}
	\includegraphics[width=0.45\textwidth]{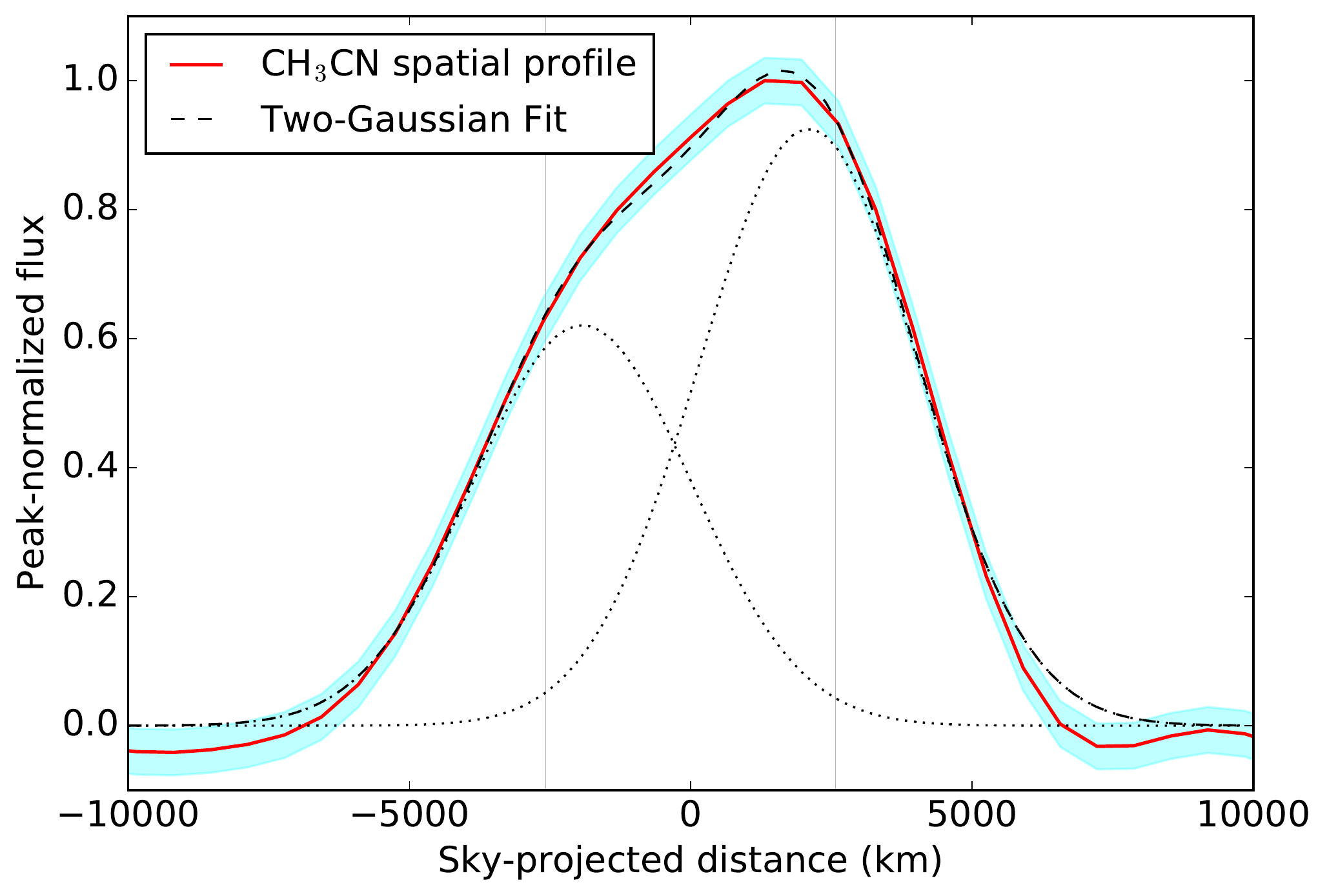}
	\caption{Peak-normalized spatial emission profiles taken along the polar axes of the maps in Figure \ref{fig:maps} (from south to north). Results of the two-component Gaussian fits (parameters in Table \ref{tab:gauss}) are overlaid. Colored envelopes show the $1\sigma$ error margins for each species. Vertical gray lines indicate the south and north polar limbs, respectively. \label{fig:profiles}}
\end{figure*}

\section{Discussion}

\subsection{Vertical Mixing Ratio Profiles}
The detected \vycn\ transitions are optically thin and their strengths exhibit a weak temperature dependence in the range ($\approx160-180$~K) found in Titan's atmosphere above 300~km. The vertical abundance profile for this species therefore implies that \vycn\ is present with an average abundance of $3.03\pm0.29$~ppb in Titan's atmosphere above 300~km. This value is in agreement, within error, with the abundance of $2.83\pm0.24$~ppb derived by \citet{pal17}. Both observations confirm substantial production of \vycn\ in Titan's upper/middle atmosphere, as predicted by the model of \citet{loi15}.  Given the downward mixing of photochemical products over time, the relatively low tropospheric \vycn\ abundance implies a short chemical lifetime for this molecule compared to those of more stable species such as C$_2$H$_2$, C$_2$H$_6$, HCN, and CH$_3$CN, which remain abundant as they travel into the lower stratosphere.

Our retrieved abundance for \etcn\ using a 300~km step model ($7.37\pm0.32$ ppb) is close to the values obtained in the previous work of \citet{cor15} and \citet{pal17} (9.25 ppb and 7.2 ppb, respectively). The C$_2$H$_5$CN/\vycn\ abundance ratio of $2.43\pm0.26$ is also consistent with the value of $2.5\pm0.3$ found by \citet{pal17}. Overall, the latest chemical models (see \citealt{loi15} and \citealt{dob16}), do not fit the \etcn\ results well, but provide a better fit for \vycn. Our newly determined \etcn\ abundance confirms a tendency for these models to over-estimate the stratospheric \etcn\ abundance. This could support the hypothesis that additional (or more rapid) destruction reactions need to be added to chemical networks. In particular, \citet{dob16} suggested that although the inclusion of ion chemistry reduced their over-estimation of the \etcn\ abundance, the remaining over-estimation is suggestive of a destruction mechanism which still remains unaccounted for. Alternatively, the discrepancy could be due to the fact that no current photochemical models have incorporated two-dimensional dynamics yet. Further modeling will be required to elucidate whether and which of these changes could lead to improved modeled abundances for \etcn\ in Titan's atmosphere.

Our disk-averaged ALMA abundance profile for CH$_3$CN can be compared with that obtained by \citet{mar02}, based on observations using the IRAM 30-m telescope. Our abundance profiles are generally in good agreement, except in the lower regions of the atmosphere. Below 200~km, the abundance profile of \citet{mar02} drops off much more rapidly with altitude (see their Figure 12). This could be a manifestation of stronger downward mixing of photochemical products around the time of our observations, which could be explained as a result of the closer temporal proximity to Titan's 2017 solstice, resulting in stronger atmospheric downwelling over the winter pole. Comparing our results to the models of \citet{loi15} and \citet{dob16}, the modeled CH$_3$CN abundances increase more rapidly with altitude than the values we retrieve. As in the case of \etcn, this discrepancy may indicate that a destruction mechanism remains unaccounted for in the lower atmosphere. \citet{dob16} suggest that these processes may include meridional transport or sticking to aerosol grains. There are also significant differences between our retrieved CH$_3$CN profile and the model of \citet{kra09}, with an observed abundance at 500~km of $6\times10^{-8}$ compared to the modeled value of $1\times10^{-6}$ (see his Figure 9). This suggests that the chemical production and destruction scheme for this molecule on Titan still remains to be fully understood.

In comparison to previously reported results for HC$_3$N \citep{mar02,tea07,vin10,cor14}, our retrieved abundance for this molecule is greater in the lower stratosphere (at altitudes up to about 300~km). Above this level, our retrieved profile matches previous profiles well, with the exception of that retrieved by \citet{cor14}, who obtained a profile with a lower abundance of HC$_3$N above this altitude. When compared to models by \citet{loi15}, our retrieved profile falls within their 90\% uncertainty interval. Given this good agreement with modeling results, it would appear that our results support the proposed formation of HC$_3$N via the reaction of CN with C$_2$H$_2$. Nonetheless, \citet{dob16} suggest that this molecule's abundance may also be affected by the aforementioned missing destruction mechanisms; if this were accounted for, modeled abundances in the lower troposphere may in fact match our retrieved abundances with an even better fit.

\subsection{Molecular Emission Maps}
\label{sec:maps}
The additional \vycn\ emission lines detected not only increase our confidence in the previous detection by \citet{pal17}, but have allowed the first mapping of the spatial distribution of this molecule. Our \vycn\ map (Figure \ref{fig:maps}) shows a flux peak at $5\sigma$ confidence in Titan's southern hemisphere, consistent with enrichment over the south polar region. A similar south-polar enrichment is also evident for HC$_3$N and C$_2$H$_5$CN. Polar enrichments have previously been observed for a number of hydrocarbons and nitriles by Cassini and ALMA \citep{tea13,cor15,vin15}. It has been theorized that as Titan transitions from northern winter in 2004 toward the southern winter solstice in 2017 May, the main atmospheric circulation cell, responsible for the redistribution of photochemical products from mid-latitudes toward the poles, should reverse its direction \citep[\eg][]{tea12}. The abundances of these molecules are thus expected to begin increasing in the southern polar region, while molecules concentrated in the north polar region would begin to undergo photochemical destruction, with a corresponding decay in the previous northern peak at a rate in accordance with the photochemical lifetime and diffusion rate of each molecule.

A clearer view of the latitudinal variability of the observed gases is shown in Figure \ref{fig:profiles}, where we present the peak-normalized emission line fluxes as a function of sky-projected distance along Titan's polar axis from south to north.  Two-component Gaussian fits with freely variable positions and intensities for the Gaussian components have been performed to these profiles, but with FWHM fixed at the instrumental resolution of 4600~km. These fits are overlaid in Figure \ref{fig:profiles} with dashed lines, and the positions and relative strengths of the components are given in Table \ref{tab:gauss}. For all species, a reasonably good fit was obtained within the noise. The positions of the fit components relative to Titan's 2575~km radius are consistent with the majority of the emission originating from near the poles, and from within a region much smaller than the spatial resolution element. Such confinement, whereby the majority of the gas is located within $20^{\circ}$ of the pole, has previously been observed by Cassini for HC$_3$N and other short-lived species \citep[\eg][]{tea08}. Discrepancies between the fits and observations may be due to noise or to the presence of additional flux components.

The one-dimensional spatial flux profiles highlight the pronounced southern peak for HC$_3$N (with a contrast factor of about three relative to the northern peak).  The factor of $\sim$3 southern enhancement of the \vycn\ abundance appears to be similar to that of HC$_3$N, whereas the somewhat smaller factor of $\sim$1.5 enhancement in \etcn\ implies either a slower production rate for \etcn\ in the south, or a slower destruction rate in the north compared with HC$_3$N and \vycn. This result is in contrast to the ALMA observations from 2012 by \citet{cor15}, which showed C$_2$H$_5$CN to be enriched in the south while HC$_3$N was enriched in the north, leading to the conclusion that C$_2$H$_5$CN has the shorter chemical lifetime. Only three terrestrial years later, C$_2$H$_5$CN remains concentrated in the south, while the location of the HC$_3$N peak appears to have shifted from the north to the south. However, \citet{cor15} did not present information on the vertical distribution of HC$_3$N, so it is possible that the northern HC$_3$N peak in that study is caused by the presence of colder, higher-pressure gas, at lower altitudes than that probed by the vibrationally excited HC$_3$N lines in our study. Continued spectral, spatial, and temporal monitoring of these molecules will be required to further elucidate the details of their relative production, destruction, and transport rates as a function of altitude. 

Significant stratospheric temperature differences have previously been identified between Titan's northern and southern polar regions, which could cause differences in the emission line strengths as a result of (1) the temperature dependence of the molecular excitation, (2) the Planck radiation function, and (3) the molecular partition functions.  To assess the possible impact of these temperature effects, we extracted spatially resolved CH$_4$ temperature data for the complete latitudinal range from the T102, T105, T110, and T111 Cassini flybys (between 2014 June and 2015 April), as explained in Section \ref{sec:results}. These temperature data were averaged within $0.25''$ Gaussian beams centered over the north and south polar limbs, chosen to cover the approximate extent of the polar latitudinal regions within which our observed molecules are likely to be concentrated \citep[see][]{tea08}. Between $z=250$-$400$~km, the temperature difference between the north and south limbs was found to be 5-8~K (see Figure \ref{fig:temps}), and at all other altitudes, the temperature difference was less than 5~K. The resulting temperature dependence of our detected emission line strengths was found to be $<20$\% for HC$_3$N, $<10$\% for CH$_3$CN, and negligible for the other species. Combined with the fact that the observed emission is quite optically thin, the structure in the spatial flux profiles can thus be interpreted as primarily due to latitudinal variations in the observed gas abundances. 

In contrast to the other species observed in the present study, the CH$_3$CN flux peaks in Titan's northern hemisphere, which is consistent with a greater concentration of this species over the north pole. Given the near-complete transition to northern summer following the northern spring equinox in 2009, this remnant gas concentration indicates that CH$_3$CN is more resistant to photochemical destruction than the other species observed, which agrees with the relatively long ($\approx8$~year) theorized photochemical lifetime for CH$_3$CN compared to the shorter ($\lesssim1$~year) lifetimes for \vycn, \etcn, and HC$_3$N \citep[see, \eg,][]{kra09}. Our latitudinal profiles are qualitatively consistent with the model results of \citep{loi15}, who determined longer lifetimes for CH$_3$CN and \etcn\ than for HC$_3$N and \vycn. The differences in these molecules' lifetimes in Titan's atmosphere may be explained by their degree of saturation: CH$_3$CN and \etcn\ are fully saturated in the sense that all carbon atoms are bonded to four other atoms (with the exception of the carbon in the nitrile group), whereas HC$_3$N and \vycn\ are unsaturated, containing carbon atoms that are double- or triple-bonded. Saturated molecules are less reactive, which explains the longer lifetimes of CH$_3$CN and \etcn\ relative to HC$_3$N and \vycn. Furthermore, CH$_3$CN and \etcn\ are less rapidly photodissociated than HC$_3$N and \vycn\, due to line shielding by CH$_4$ (J. C. Loison 2017, private communication). A similar relationship in chemical lifetimes between HC$_3$N, \etcn, and CH$_3$CN is produced in the latest chemical model of V. Vuitton (2017, private communication). This model also points to photolysis rates as an important factor in the chemical lifetimes for these species. In particular, HC$_3$N is more efficiently photolysed than CH$_3$CN by the longer wavelength photons found in the stratosphere.

The importance of Titan's polar tilt with respect to the line of sight (currently $24.7^{\circ}$) needs to be considered when interpreting latitudinal flux profiles. Whereas Titan's north polar surface was in full view of Earth at the epoch of our observations, the region south of $-65^{\circ}$ was obscured behind Titan's disk. The detailed distribution of Titan's CH$_3$CN is currently not well-known, but if a significant proportion is concentrated at low altitudes over the pole, this could be responsible for some of the apparent northerly enhancement. Conversely, the southerly enhancement for the other observed species is likely to be even more pronounced if the south pole were in full view.

\section{Conclusion}
We have presented here observations of three new transitions of vinyl cyanide, \vycn, strengthening our confidence in the previously reported spectroscopic detection of this molecule by \citet{pal17}. We also report a tentative spectroscopic detection of propane, C$_3$H$_8$.

Abundances were derived for \vycn, \etcn, HC$_3$N, and CH$_3$CN using radiative transfer modeling. These results show a generally a good agreement with prior observations. The retrieved abundance for \vycn\ is $3.03\pm0.29$~ppb above 300~km, and our models support previous indications that this molecule is produced efficiently on Titan at high altitudes. Discrepancies between our retrieved abundance profiles and chemical models for CH$_3$CN, \vycn, and \etcn\ highlight the need for improved understanding of the chemistry of these species in planetary atmospheres. 

We were able to produce the first spatially resolved map of the distribution of \vycn\ on Titan, as well as new maps of \etcn, HC$_3$N, and CH$_3$CN for the 2015 April epoch. Comparing these to maps produced in prior epochs, we find support for a global redistribution of atmospheric nitriles, in accordance with the hypothesized seasonally driven global circulation cell direction change. Furthermore, the observed changes lend support to chemical model predictions that HC$_3$N and \vycn\ have a shorter lifespan in Titan's atmosphere than \etcn\ and CH$_3$CN. The continued northern enrichment of CH$_3$CN despite the change in season supports a relatively longer chemical lifespan than for the other observed species. 

Following our detection and mapping of \vycn\ and other nitriles using ALMA, additional detailed mapping (and chemical modeling) of the time-variability of these species is warranted to further understand their photochemistry in primitive planetary atmospheres.

\acknowledgments

This work was supported by the NASA Astrobiology Institute through funding awarded to the Goddard Center for Astrobiology under proposal 13-13NAI7-0032, and by the National Science Foundation under Grant No. AST-1616306. It makes use of ALMA data set ADS/JAO.ALMA\#2013.1.00033.S. ALMA is a partnership of ESO (representing its member states), NSF (USA), and NINS (Japan), together with NRC (Canada), and NSC and ASIAA (Taiwan), in cooperation with the Republic of Chile. The Joint ALMA Observatory is operated by ESO, AUI/NRAO, and NAOJ. The National Radio Astronomy Observatory is a facility of the National Science Foundation operated under cooperative agreement by Associated Universities, Inc. C.A.N. and S.B.C. were supported by the NASA HQ Science Innovation Fund for their portion of the work in this paper.

\emph{Software}: CASA v.4.5.3 \citep{mcm07}, NEMESIS \citep{irw08}

\clearpage

\end{document}